\documentclass[prb,aps,12pt]{revtex4-2}
\usepackage{amsmath}
\usepackage{amssymb}
\usepackage{graphicx}

\begin{document}

\title{Frustrated spin models on two- and three-dimensional
decorated lattices with high residual entropy}
\author{D.~V.~Dmitriev}
\email{dmitriev@deom.chph.ras.ru}
\author{V.~Ya.~Krivnov}
\author{O.~A.~Vasilyev}
\affiliation{Institute of Biochemical Physics of RAS, Kosygin str.
4, 119334, Moscow, Russia.}
\date{}

\begin{abstract}

We study the ground-state properties of a family of frustrated
spin-1/2 Heisenberg models on two- and three-dimensional decorated
lattices composed of connected star-shaped units. Each star is
built from edge-sharing triangles with an antiferromagnetic
interaction on the shared side and ferromagnetic interactions on
the others. At a critical coupling ratio, the ideal star model
- defined by equal ferromagnetic interactions - exhibits a
macroscopically degenerate ground state, which we map
onto a site percolation problem on the Lieb lattice. This mapping
enables the calculation of exponential ground-state degeneracy
and the corresponding residual entropy for square, triangular,
honeycomb, and cubic lattices. Remarkably, the residual entropy 
remains high for all studied lattices, exceeding 60\% of the  
maximal value ln(2). Despite a gapless quadratic
one-magnon spectrum, the low-temperature thermodynamics is
governed by exponentially numerous gapped excitations. 
For a distorted-star variant of the model, the ground-state manifold 
is equivalent to that of decoupled ferromagnetic clusters, leading 
to exponential degeneracy with a lower, yet still substantial, 
residual entropy. At low temperature the system mimics a paramagnetic 
crystal of non-interacting spins with high spin value 
($s=4$ for a square lattice). The obtained results establish a structural 
design principle for engineering quantum magnets with a high 
ground-state degeneracy, suggesting promising candidates for
enhanced magnetocaloric cooling and quantum thermal machines.

\end{abstract}

\maketitle

\section{Introduction}

Quantum magnets on geometrically frustrated lattices often host
macroscopically degenerate ground states, a phenomenon closely
tied to flat, dispersionless magnon bands \cite{diep, diep2,
mila}. This flat-band physics, highlighted in seminal studies of
highly frustrated antiferromagnets
\cite{schulenburg2002macroscopic, flat}, arises from localized
magnon states confined to specific lattice cells by destructive
interference \cite{derzhko2007universal}. These localized states
can be combined into exact multi-magnon eigenstates, enabling a
mapping to a classical lattice gas model \cite{flat, zhit,
strevcka2017diversity}. Such degeneracy leads to distinctive
thermodynamic features like magnetization plateaus and enhanced
magnetocaloric effects \cite{flat,derzhko2007universal,shulen,
richter2018thermodynamic, zhitomirsky2003enhanced}.

A distinct route to high degeneracy emerges in systems with
competing ferromagnetic (F) and antiferromagnetic (AF)
interactions. At a quantum critical point where different phases
meet, these F-AF models, such as the frustrated delta-chain and
its extensions \cite{zhitomir, Derzhko, KDNDR, DKRS, DKRS2, anis,
anis2, anis3}, exhibit macroscopic degeneracy and high residual
entropy even at zero field - a property relevant for magnetic
cooling applications. Beyond adiabatic demagnetization, such
macroscopic degeneracy at a quantum critical point is also central
to the operation of quantum thermal machines. Enhanced work output
and efficiency can be achieved when a thermodynamic cycle is
operated across a quantum critical point, utilizing the large
ground-state degeneracy $g_{\max }$ as an 'entropic lever'. The
maximal net work extractable scales as $W=k_{B}\left( T_{\max
}-T_{\min }\right) \ln \left( g_{\max }\right) $, directly linking
the performance of the machine to the extensive degeneracy of the
model \cite{PhysRevE.96.022143,
purkait2022performance,castorene2024effects}.

Recent work has extended this framework to models built from
diamond-shaped clusters. A spin-$\frac{1}{2}$ chain of distorted
diamonds was shown to host flat bands for one-, two-, and
three-magnon states \cite{diamond1d}. Similarly, models on
diamond-decorated square and cubic lattices support up to five or
seven localized magnons per cell \cite{DKV2D}. The inclusion of
these multi-magnon states substantially amplifies ground-state
degeneracy. Experimentally, diamond units with competing F and AF
interactions are realized in materials such as
\mbox{$K_{3}Cu_{3}AlO_{2}(SO_{4}){4}$} \cite{alum} and
\mbox{$K_{2}Cu_{3}(MoO_{4})_{4}$} \cite{PhysRevB.111.144420}.

Purely antiferromagnetic models of ideal diamonds also exhibit
rich physics, with phases including macroscopically degenerate
monomer-dimer states \cite{takano, morita2016exact,
hirose2016exact, caci2023phases}. However, a different degeneracy
mechanism operates in F-AF models with ideal diamonds at a special
coupling ratio. Here, the ground state factorizes into decoupled
clusters with identical energy per spin, leading to exponential
degeneracy from isolated ferromagnetic clusters embedded in a
singlet background \cite{diamond1d, DKV2D}.

In this work, we explore a structural variant of these models:
lattices constructed from connected star units rather than
conventional bond decorations. We demonstrate that this
architecture sustains exceptionally high and robust residual
entropy across various two- and three-dimensional lattices,
highlighting its potential as a design principle for high-entropy
quantum magnets and, consequently, for applications in efficient
cooling and quantum thermal machines.

The remainder of this paper is organized as follows. In Section
II, we introduce the two- and three-dimensional lattices
constructed from edge-sharing triangular units. We demonstrate
that the Hamiltonian for so constructed models maps onto a Lieb
lattice, with composite spins $L=0$ (singlet) or $L=1$ (triplet)
occupying its sites. This mapping allows us to calculate the
exponential ground-state degeneracy and residual entropy via a
connection to a classical site percolation problem. Section III
explores the same models with distorted unit blocks, analyzing how
this modification influences the ground-state degeneracy and
residual entropy. Finally, in Section IV, we summarize our key
findings.

\section{Ideal star model on the 2D and 3D lattices}

\begin{figure}[tbp]
\includegraphics[width=6in,angle=0]{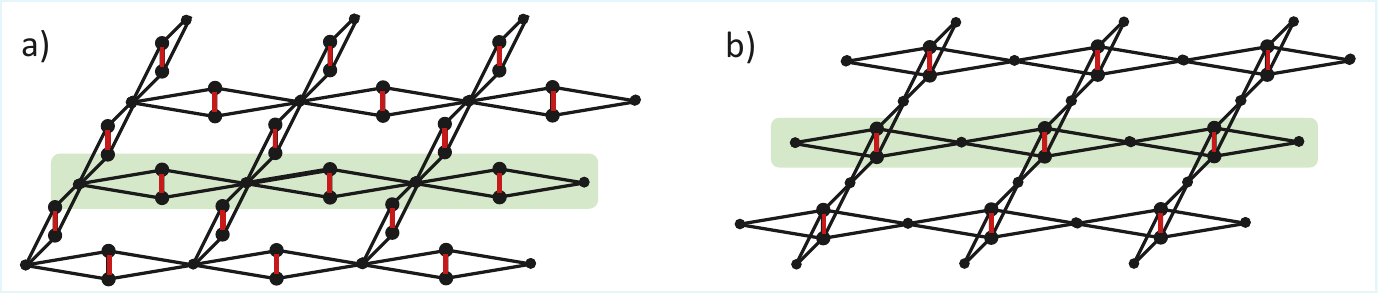}
\caption{Two versions of the extension of the diamond chain
(shaded) to the square lattice: a) the diamond-decorated square
lattice, and b) the square lattice formed by connected stars.}
\label{Fig_square}
\end{figure}

In our previous work \cite{DKV2D}, we introduced an extension of
the diamond chain model with high ground state degeneracy
\cite{diamond1d} to two- and three-dimensional diamond-decorated
lattices. Here, we consider an alternative structural variant, in
which the diamond diagonals occupy the lattice sites, rather than
the central spins used in \cite{DKV2D}. Figure \ref{Fig_square}
compares these two extensions for the square lattice: panel (a)
shows the diamond-decorated lattice studied in \cite{DKV2D}, while
panel (b) illustrates the present alternative construction.
Although we explicitly formulate the extension for a square
lattice, the analysis and results are fully general and apply to
any two- or three-dimensional lattice.

\begin{figure}[tbp]
\includegraphics[width=4in,angle=0]{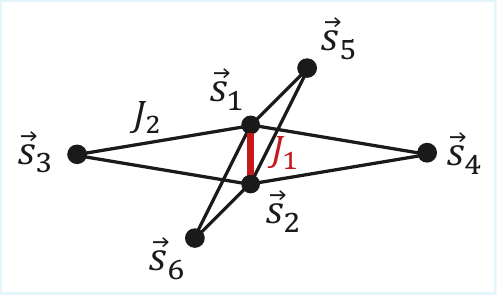}
\caption{An elementary cell (star) for the square lattice.}
\label{Fig_star}
\end{figure}

As shown in Fig. \ref{Fig_square}b, the square lattice is
constructed from connected elementary cells, or stars. A single
star is illustrated in Fig. \ref{Fig_star}. It consists of
edge-sharing triangles with an antiferromagnetic exchange $J_1$
(red) on the shared side and equal ferromagnetic exchanges $J_2$
(black) on the remaining sides. In the following, we refer to the
spins on the diagonal as dimer spins and the outer spins as
monomer spins.

The total Hamiltonian is a sum over all local star Hamiltonians:
\begin{equation}
\hat{H}=\sum_{\mathbf{i}}\hat{H}_{\mathbf{i}}  \label{Hh}
\end{equation}

The local Hamiltonian for the single star shown in
Fig.\ref{Fig_star} has the form
\begin{equation}
\hat{H}_{0}=J_{1}\mathbf{s}_{1}\cdot \mathbf{s}_{2}-J_{2}\left(
\mathbf{s}_{1}\mathbf{+s}_{2}\right) \cdot \left(
\mathbf{s}_{3}\mathbf{+s}_{4}\mathbf{+s}_{5}\mathbf{+s}_{6}\right)
+\frac{3}{4}J_{1}  \label{Hstar}
\end{equation}%
where $\mathbf{s}_{i}$ are spin-$\frac{1}{2}$ operators and the
constant in (\ref{Hh}) is added for convenience, it ensures the
ground state energy of the singlet sector is zero.

As follows from Hamiltonian (\ref{Hstar}), there is a local
conservation of a composite spin
$\mathbf{L}=\mathbf{s}_{1}+\mathbf{s}_{2}$ on the diagonal of the
star (dimer). Composite spin is a conserved quantity with a
quantum spin number $L=0$ or $L=1$, which corresponds to the
singlet or triplet state of the dimer, respectively. The singlet
state of the dimer shown in Fig.\ref{Fig_star}, $\left(
s_{1}^{-}-s_{2}^{-}\right) \left\vert F\right\rangle $ (where
$\left\vert F\right\rangle $ is the fully polarized state with all
spins down), is an exact state of $\hat{H}_{0}$, independent of
the configuration of the monomer spins
$\mathbf{s}_{3},\mathbf{s}_{4},\mathbf{s}_{5},\mathbf{s}_{6}$. In
terms of the composite spin, the local Hamiltonian (\ref{Hstar})
simplifies to
\begin{equation}
\hat{H}_{0}=\frac{J_{1}}{2}\mathbf{L}^{2}-J_{2}\mathbf{L}\cdot \left(
\mathbf{s}_{3}\mathbf{+s}_{4}\mathbf{+s}_{5}\mathbf{+s}_{6}\right)
\label{Hstar2}
\end{equation}

The ground state of (\ref{Hstar2}) depends on the relation between
the exchanges $J_{1}$ and $J_{2}$. For $J_{1}<2J_{2}$, the
ferromagnetic state with $L=1$ and the total spin $S_{tot}=3$ is
the ground state of $\hat{H}_{0}$, with the energy
$E_{F}=J_{1}-2J_{2}$. For $J_{1}>2J_{2}$, the ground state of
$\hat{H}_{0}$ is degenerate and has $L=0$, which means that the
four monomer spins $\mathbf{s}_{3},\ldots \mathbf{s}_{6}$ are
effectively free. This yields a local ground-state manifold
composed of one quintet ($S_{tot}=2$), three triplets
($S_{tot}=1$), and two singlets ($S_{tot}=0$), all with energy
$E_{0}=0$.

Therefore, the global ground state phase diagram of model
(\ref{Hh}) consists of the ferromagnetic (F) for $J_{1}<2J_{2}$
and monomer-dimer (MD) phases  for $J_{1}>2J_{2}$. The F phase,
with all $L_{i}=1$, has a degeneracy of $(\mathcal{N}+1)$, where
$\mathcal{N}$ is the total number of spins in the system. The
monomer-dimer (MD) phase, with all $L_{i}=0$, has a macroscopic
ground state degeneracy of $2^{N_{bond}}$, where $N_{bond}$ is the
number of free monomer spins $\mathbf{s}_{\mathbf{i},\mathbf{j}}$
located on the bonds between the sites $\mathbf{i}$ and
$\mathbf{j}$.

At the quantum critical point $J_{1}=2J_{2}$, the ferromagnetic
state of local Hamiltonian (\ref{Hstar2}) with $L=1$ and
$S_{tot}=3$ becomes degenerate with the $L=0$ ground state
manifold. This additional degeneracy of the local Hamiltonian at
$J_{1}=2J_{2}$ induces a significantly higher ground state
degeneracy of the total system. In this work, we focus on this
quantum critical point, which separates the F and MD phases. For
convenience, we set the energy scale with $J_{2}=1$ and take
$J_{1}=2$.

\begin{figure}[tbp]
\includegraphics[width=3in,angle=0]{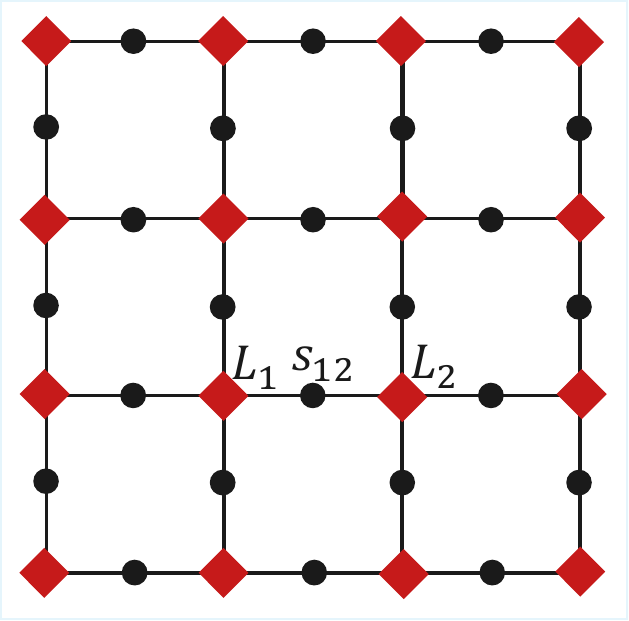}
\caption{Mapping of the model to a square Lieb lattice. Sites host
composite spins $\mathbf{L}_{\mathbf{i}}$; bonds host monomer
spins-$\frac{1}{2}$, $\mathbf{s}_{\mathbf{i},\mathbf{j}}$.}
\label{Fig_Lieb_square}
\end{figure}

Consequently, the local Hamiltonians $\hat{H}_{\mathbf{i}}$ in
(\ref{Hh}) centered at site $\mathbf{i}=\left( i_{x},i_{y}\right)$
of the square lattice can be expressed as
\begin{equation}
\hat{H}_{\mathbf{i}}=\mathbf{L}_{\mathbf{i}}^{2}-\sum_{\mathbf{j}}\mathbf{L}
_{\mathbf{i}}\cdot \mathbf{s}_{\mathbf{i},\mathbf{j}}  \label{Hd}
\end{equation}%
where $\mathbf{s}_{\mathbf{i},\mathbf{j}}$ are spin-$\frac{1}{2}$
operators on the bonds between sites $\mathbf{i}$ and
$\mathbf{j}$. As illustrated in Fig.\ref{Fig_Lieb_square}, the
original spin-$\frac{1}{2}$ system maps onto a Lieb lattice: the
monomer spins $\mathbf{s}_{\mathbf{i},\mathbf{j}}$ reside on
bonds, and the composite spins $\mathbf{L}_{\mathbf{i}}$ occupy
the sites of the Lieb lattice.

For an arbitrary lattice with coordination number $z$, the general
Hamiltonian can be written in the form
\begin{equation}
\hat{H}=\frac{z}{4}\sum_{\mathbf{i}}\mathbf{L}_{\mathbf{i}%
}^{2}-\sum_{\left\langle \mathbf{i,j}\right\rangle }\left(
\mathbf{L}_{\mathbf{i}}+\mathbf{L}_{\mathbf{j}}\right) \cdot
\mathbf{s}_{\mathbf{i},\mathbf{j}}  \label{H}
\end{equation}%
where the second sum runs over all nearest-neighbor sites
$\mathbf{i}$ and $\mathbf{j}$ on the corresponding lattice.

\subsection{Ground state degeneracy}

A singlet state of a dimer at site $\mathbf{i}$
($L_{\mathbf{i}}=0$) effectively decouples the neighboring bond
spins $\mathbf{s}_{\mathbf{i},\mathbf{j}}$. Therefore it is more
convenient to compute the ground state degeneracy not in terms of
magnon sectors (or the total $S^{z}$), but by enumerating all
possible configurations of such singlet sites on the Lieb lattice.
Each ground state is uniquely associated with a specific
configuration of singlet sites; the total degeneracy is obtained
by summing contributions from every such configuration.

Thus, the problem maps exactly onto a \textit{site percolation
problem on the Lieb lattice}, where sites in the triplet state
($L_{\mathbf{i}}=1$) are considered `connected' and singlet sites
($L_{\mathbf{i}}=0$) are `disconnected'. The calculation follows
the procedure of Ref.\cite{DKV2D}, which addressed a \textit{bond
percolation} problem for diamond-decorated lattices; here we adapt
it to site percolation.

For a particular configuration $\omega _{K}$ with $K$ connected
sites (triplet states), the lattice effectively decomposes into
disconnected ferromagnetic clusters. For the $i$-th cluster
containing $n_{i}$ connected sites and $l_{i}$ bonds, the total
number of spins-$\frac{1}{2}$ is $(2n_{i}+l_{i})$, yielding a
cluster degeneracy $(2n_{i}+l_{i}+1)$ (all possible $S^{z}$
projections of ferromagnetic ground state). The total degeneracy
for configuration $\omega _{K}$ is the product of the numbers of
ground states over all clusters
\begin{equation}
W(\omega _{K},N)=\prod_{i\in \omega _{K}}(2n_{i}+l_{i}+1)  \label{W}
\end{equation}

\begin{figure}[tbp]
\includegraphics[width=4in,angle=0]{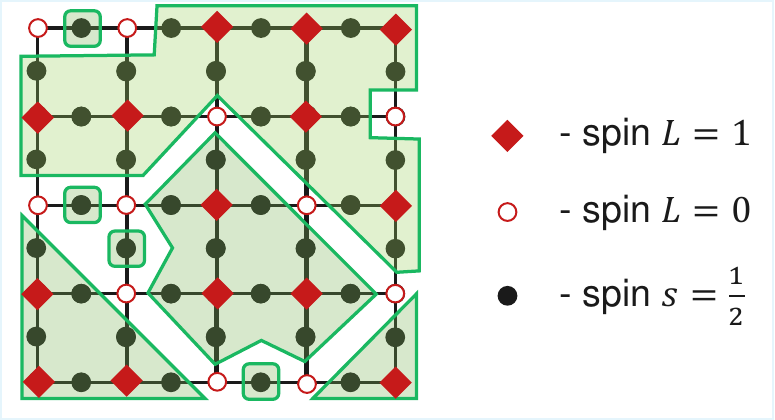}
\caption{A 4x4 Lieb lattice with a specific configuration of
singlet sites (open circles, $L_i=0$). Shaded regions denote the
resulting ferromagnetic clusters.} \label{Fig_percolation}
\end{figure}

As an illustration, Fig.\ref{Fig_percolation} shows a specific
configuration of `connected' sites on $4\times 4$ square lattice
with open boundaries. In this configuration, eight distinct
clusters emerge: four single-spin clusters and clusters containing
$4,12,16,32$ spins-$\frac{1}{2}$. According to Eq.(\ref{W}), the
ground state degeneracy for this particular configuration is
$W=2^{4}\cdot 5\cdot 13\cdot 17\cdot 33=583440$ states.

The total degeneracy for a fixed number $K$ of connected sites on
an $N$-site lattice is obtained by summing over all possible
configurations $\omega _{K}$
\begin{equation}
W(K,N)=\sum_{\omega _{K}}W(\omega _{K},N)  \label{ZK}
\end{equation}

\begin{figure}[tbp]
\includegraphics[width=5in,angle=0]{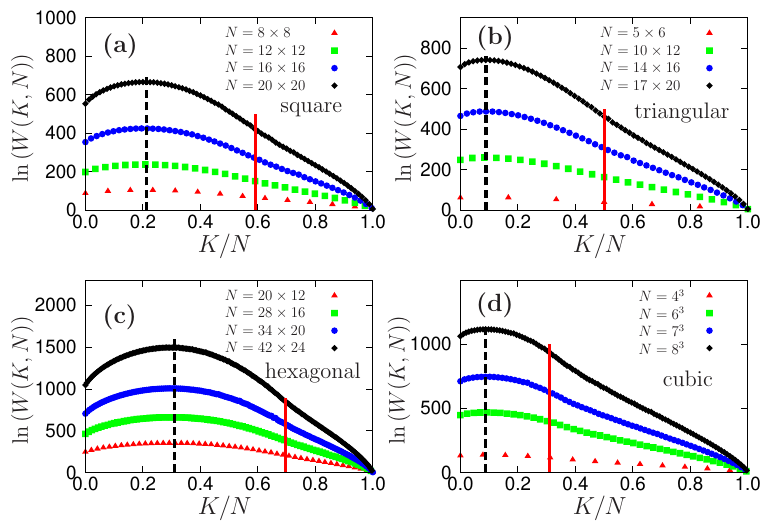}
\caption{Contributions to partition function $\ln W(K,N)$
(Eq.(\ref{ZK})) as a function of the connected-site fraction
$p=K/N$ and different system sizes $N$ and lattices: (a) square;
(b) triangular; (c) hexagonal; (d) cubic. Vertical red lines
indicate the corresponding site-percolation thresholds.}
\label{fig:ZB}
\end{figure}

\begin{figure}[tbp]
\includegraphics[width=5in,angle=0]{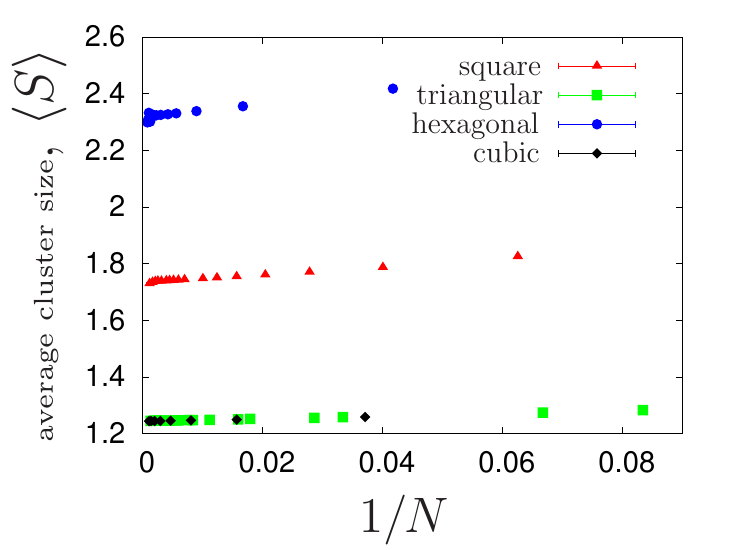}
\caption{Average cluster size $\langle S\rangle $ as a function of
the $1/N$ for different lattices.} \label{fig:Scl}
\end{figure}

Figure \ref{fig:ZB}(a)--(d) plots $\ln W(K,N)$ versus the fraction
of connected sites $p=K/N$. For all lattices, $W(K,N)$ peaks well
below the corresponding site-percolation threshold $p_c$ (red
vertical lines), at approximately $p_{0}\simeq 0.3 $ (hexagonal),
$p_{0}\simeq 0.2$ (square), $p_{0}\simeq 0.1$ (triangular and
cubic) lattices. Below $p_c$, the system consists of many small
clusters whose number scales with $N$, leading to exponential
scaling $W\sim \exp (const\cdot N)$. Above $p_c$, a single
infinite ferromagnetic cluster dominates, and $W$ scales only
linearly with $N$. Since $p_{0}<p_{c}$ for all studied lattices,
the dominant contributions to the total degeneracy come from
configurations of small, finite clusters. This is confirmed by the
average cluster size $\langle S\rangle $ shown in
Fig.\ref{fig:Scl}, which are really small (for triangular and
cubic lattices most of clusters contain just one spin) and tends
to a constant as $N\to\infty$.

The total number of ground states is given by the sum over all
possible numbers of connected bonds $K$
\begin{equation}
W(N)=\sum_{K=0}^{N}W(K,N)  \label{Z}
\end{equation}

Using the numerical Monte-Carlo approach details in
Ref.\cite{DKV2D}, we find that for all studied lattices
(hexagonal, square, triangular, cubic), the ground state
degeneracy grows exponentially with $N$
\begin{equation}
W(N)\sim e^{\alpha N}
\end{equation}%
with a lattice-dependent exponent $\alpha $. This leads to a
residual entropy per spin
\begin{equation}
\mathcal{S}_{0}(N)=\frac{\ln (W(N))}{\mathcal{N}}  \label{res_entropy}
\end{equation}%
where $\mathcal{N}=2N+\frac{1}{2}zN$ is the total number of
spins-$\frac{1}{2}$.

\begin{figure}[tbp]
\includegraphics[width=5in,angle=0]{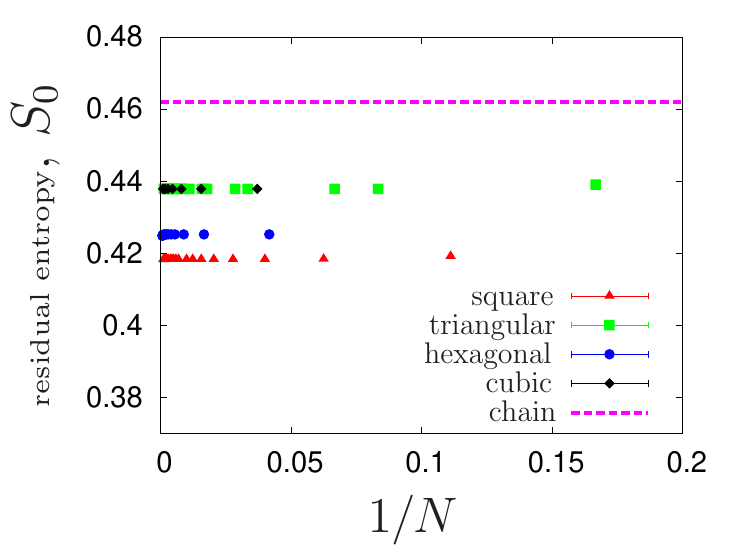}
\caption{Residual entropy $\mathcal{S}_{0}(N)$ as a function of
$1/N$ for the square, triangular, honeycomb and cubic lattices.}
\label{fig:zn}
\end{figure}

The results for $\mathcal{S}_{0}(N)$ vs. $1/N$ for different
lattices are plotted in Fig.~\ref{fig:zn}.

Figure \ref{fig:zn} shows that $\mathcal{S}_{0}(N)$ converges to a
finite thermodynamic limit for all lattices, which gives the
thermodynamic values of the residual entropy
\begin{equation}
\mathcal{S}_{0}=\frac{\alpha }{2+\frac{z}{2}}  \label{res_entropy2}
\end{equation}

Table 1 lists these limiting values of $\mathcal{S}_{0}$,
alongside the residual entropies for diamond-decorated models
\cite{DKV2D} and for the monomer-dimer (MD) phase of the present
model.

\begin{table}[tbp]
\caption{Residual entropy per spin, $\mathcal{S}_{0}$, for
diamond-decorated model, connected star model at the quantum
critical point and connected star model in the MD phase}%
\begin{ruledtabular}
\begin{tabular}{cccc}
Lattice & Diamond-decorated  & Critical point & MD phase   \\
\hline Chain ($z=2$) & $0.462$ & $0.462$ & $0.231$ \\
Hexagonal ($z=3$) & $0.402$ & $0.425$ & $0.297$ \\
Square ($z=4$) & $0.362$ & $0.418$ & $0.347$ \\
Triangular ($z=6$) & $0.314$ & $0.438$ & $0.416$ \\
Cubic ($z=6$)&  $0.302$ & $0.438$ & $0.416$ \\
\end{tabular}
\end{ruledtabular}
\end{table}

Table 1 reveals two key findings. First, the residual entropy for
connected-star models is systematically higher than for
diamond-decorated models across all lattices. Second, while the
entropy of diamond-decorated models decreases with coordination
number $z$, the residual entropy of connected-star models remains
high (exceeding 60\% of the maximal entropy $\ln 2$) and even
increases slightly with $z$ at the quantum critical point. This
contrasts with the MD phase, where the residual entropy grows with
$z$ due to the increasing number of free bond spins - a trend that
nearly saturates to the quantum critical point value for the cubic
lattice.

\subsection{Magnetization}

For a macroscopically degenerate ground state manifold, the
zero-temperature magnetization $M$ is defined by an average over
all ground states $\left\vert \psi _{k}\right\rangle $:
\begin{equation}
M^{2}=\frac{1}{W}\sum_{k=1}^{W}\left\langle \psi _{k}\right\vert
\mathbf{S} _{tot}^{2}\left\vert \psi _{k}\right\rangle  \label{M}
\end{equation}%
where $\mathbf{S}_{tot}=\sum \mathbf{s}_{i}$ is the total spin
operator and the averaging is performed over all $W$ ground states
$\left\vert \psi _{k}\right\rangle $. In the thermodynamic limit,
this definition corresponds to the square root of the long-range
spin correlation $\left\langle \mathbf{s}_{\mathbf{i}}\cdot
\mathbf{s}_{\mathbf{j}}\right\rangle $
($|\mathbf{i}-\mathbf{j}|\to \infty $), averaged over the ground
state manifold. For a pure ferromagnet with $\mathcal{N}$
spins-$\frac{1}{2}$, Eq.(\ref{M}) correctly reproduces the
saturation magnetization per spin
$m=\frac{M}{\mathcal{N}}=\frac{1}{2}$.

We calculate $m$ using the percolation mapping, following the
approach of Ref.\cite{DKV2D}. If a configuration $\omega _{K}$
contains only small clusters (the number of which is proportional
to $N$), the total spin of each cluster is of order unity, and the
magnetization per spin vanishes as $m\sim N^{-1/2}$. In contrast,
if $\omega _{K}$ contains an infinite percolation cluster of
weight $P$ (i.e., a fraction $P$ of all spins belong to a single
connected cluster), then $m=P$.

According to percolation theory, an infinite cluster exists only
above the site-percolation threshold $p_c$, growing as $P\sim
\left( p-p_{c}\right) ^{\beta }$ for $p>p_c$, with $\beta \simeq
0.14$ for 2D and $\beta \simeq 0.4$ for 3D lattices
\cite{percolation_review,beta04_percolation}. Therefore, the
presence of magnetic order depends on whether the typical value of
$p=K/N$ in the ground state ensemble lies above $p_c$. As shown in
Fig.\ref{fig:ZB}(a)-(d), the function $W(K,N)$ - and hence the
statistical weight of configurations - peaks at values $p_0$ well
below the corresponding $p_c$ for all lattices. Consequently,
configurations containing an infinite cluster are exponentially
suppressed in the thermodynamic limit, and the ground-state
magnetization must vanish.

\begin{figure}[tbp]
\includegraphics[width=5in,angle=0]{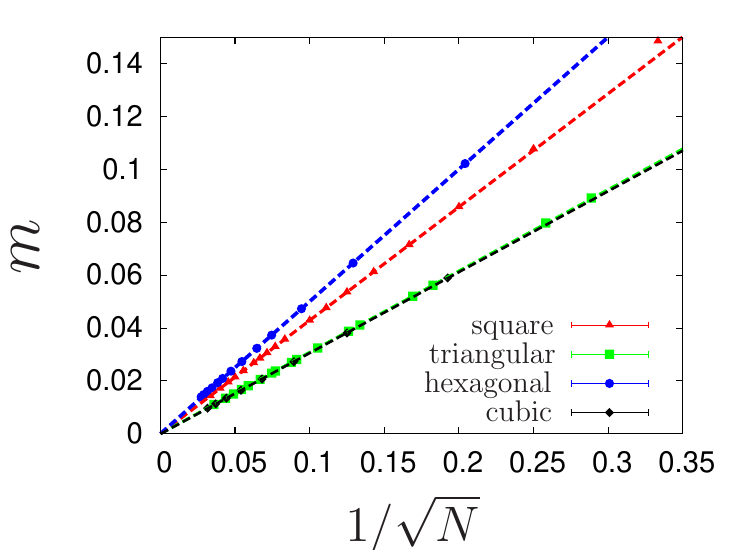}
\caption{Magnetization per spin as a function of $N^{-1/2}$ for
square, honeycomb, triangular and cubic lattices. The linear
behavior confirms $m\sim N^{-1/2}$ at large $N$.} \label{fig:M}
\end{figure}

This prediction is confirmed numerically in Fig. \ref{fig:M},
where the magnetization per spin $m(N)$ scales as $N^{-1/2}$ for
all studied lattices.

We now examine the full magnetization curve in an external
magnetic field $h$ at low temperature. In the ferromagnetic phase
($J_1<2J_2$), the magnetization behavior depends crucially on
dimensionality. For three-dimensional lattices, robust long-range
ferromagnetic order exists, leading to saturation magnetization
$m=1/2$ for an infinitesimal field. In two dimensions, however,
the Mermin-Wagner theorem precludes long-range order at finite
temperature, resulting in magnetization that increases rapidly as
$m\sim h\exp \left( const./T\right)$ before eventually saturating
at $m=1/2$.

Within the MD phase ($J_1>2J_2$), the free monomer spins on the
lattice bonds produce a linear paramagnetic response at weak
fields, $m\sim h/T$. With a further increase of the magnetic
field, all monomer spins are polarized and the magnetization comes
to a plateau at $m_{0}=\frac{z}{2z+8}$ ($m_0=1/4$ for the square
lattice). This plateau persists up to a critical field
$h_{c}=J_{1}-2J_{2}$, which corresponds to the energy gap
$E_{F}-E_{0}=J_{1}-2J_{2}$ required to excite a composite spin
from the singlet ($L=0$) to the triplet ($L=1$) state. For
$h>h_c$, all composite spins are forced into the $L=1$ state, and
the system enters the fully polarized ferromagnetic phase with
saturation magnetization $m=1/2$.

At the quantum critical point ($J_1=2J_2$), the fully polarized
state belongs to the ground state manifold at $h=0$. While an
infinitesimal field would select this state at strictly zero
temperature, the behavior at low but finite temperature $T$ is
governed by a competition between entropic and energetic
contributions. The partition function is dominated by two terms,
$Z\approx Z_{1}+Z_{2}$. The first term $Z_1$ originates from the
exponentially numerous configurations with a fraction $p\simeq
p_0$ of triplet sites below the percolation threshold $p_c$:
\begin{equation}
Z_{1}\sim \exp \left(
\mathcal{S}_{0}\mathcal{N}+\frac{h\mathcal{N}}{2T}-\frac{hN\left(
1\mathcal{-}p_{0}\right) }{T}\right)
\end{equation}

The second term $Z_2$ comes from the much fewer nearly polarized
states above the percolation threshold, which have lower Zeeman
energy. Therefore, the second term can be approximated as
\begin{equation}
Z_{2}\sim \exp \left( \frac{h\mathcal{N}}{2T}\right)
\end{equation}

Comparing $Z_{1}$ and $Z_{2}$, we find that $Z_{1}$ dominates for
$h\ll T$, whereas $Z_{2}$ prevails in the opposite limit $h\gg T$.
This means that for $h\ll T$, the system behaves as paramagnet
with $m\sim \frac{h}{T}$. When $h\gg T$, the system behaves as
ferromagnet, which means the saturation of magnetization
$m=\frac{1}{2}$.

\begin{figure}[tbp]
\includegraphics[width=5in,angle=0]{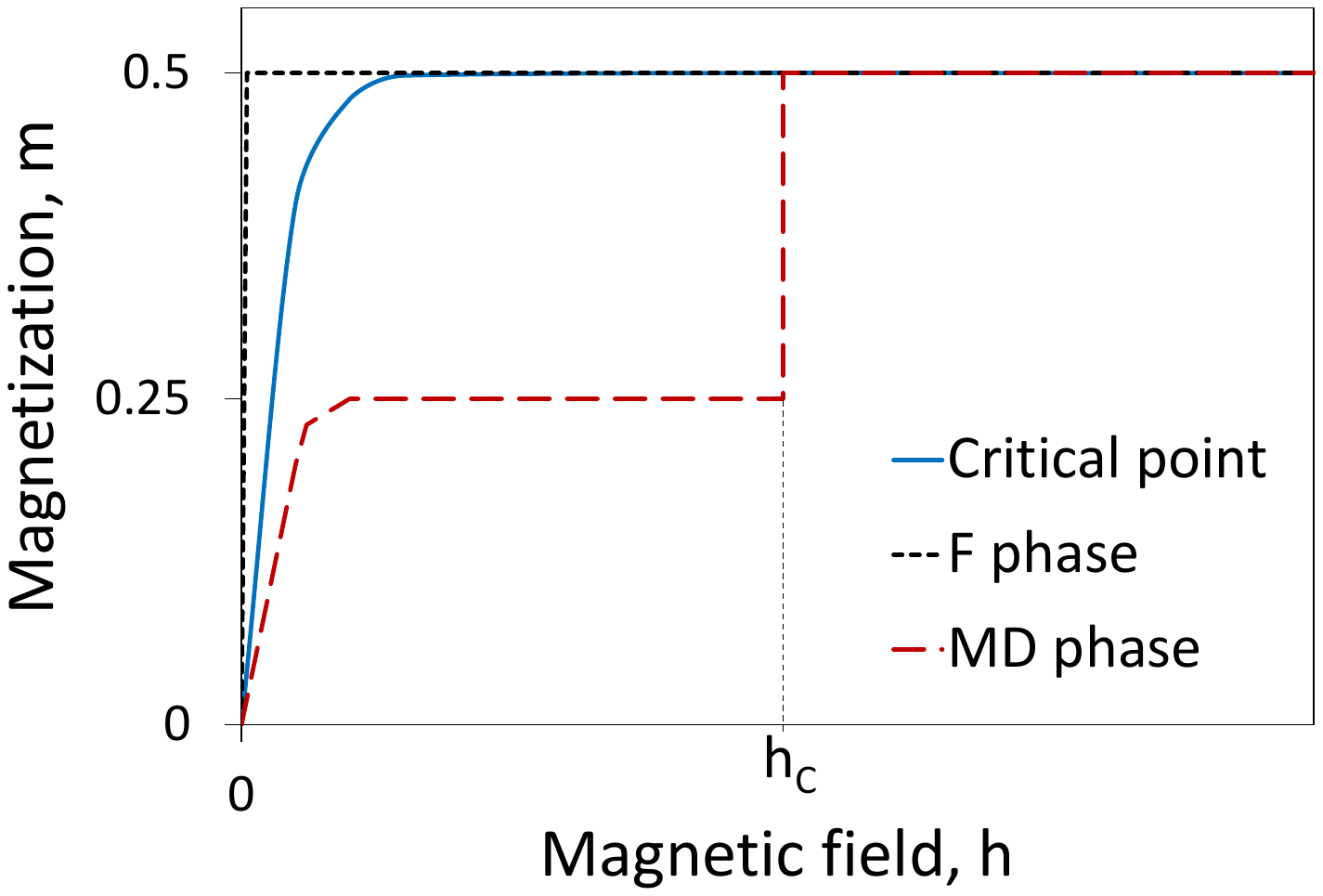}
\caption{Schematic magnetization curves in the low-temperature
limit for the ferromagnetic phase (black long-dashed line), the MD
phase (red dashed line), and at the quantum critical point (blue
solid line). In the MD phase and at the quantum critical point for
$h\ll T$, the slope is proportional to $1/T$.} \label{fig:M_h}
\end{figure}

All these results are summarized and schematically shown in
Fig.\ref{fig:M_h}.

\subsection{Excitation spectrum and specific heat}

The one-magnon excitation spectrum of the connected stars model on
the square lattice can be calculated exactly. It consists of two
flat bands with energies $E_{0}=0$ (the ground-state band) and
$E_{1}=2$, along with two dispersive branches
\begin{equation}
E_{2,3}(\mathbf{k})=2\pm 2\sqrt{1-\frac{1}{4}(2-\cos k_{x}-\cos k_{y})}
\end{equation}

Near $k_{x}=k_{y}=0$ the lower branch behaves as
\begin{equation}
E_{2}(\mathbf{k})=\frac{1}{8}\mathbf{k}^{2}  \label{k2}
\end{equation}
indicating a gapless quadratic dispersion similar to that of a
conventional ferromagnet. Consequently, on a finite system of
linear size $n$, the finite-size gap scales as $\sim n^{-2}$ or
$N^{-1}$ for $N=n^2$ sites. Analogous calculations for other
lattices (triangular, hexagonal, cubic) likewise yield a gapless,
quadratic low-energy spectrum $E(\mathbf{k} )\sim \mathbf{k}^{2}$.

Despite this gapless dispersion in the thermodynamic limit, the
dominant ground state configurations consist almost entirely of
small, disconnected ferromagnetic clusters. As established in
Sec.IIb, the statistical weight of configurations containing an
infinite (spanning) ferromagnetic cluster is exponentially
suppressed. The typical cluster size remains finite as
$N\to\infty$, implying that most local excitations are gapped.

This leads to a marked separation of energy scales in the
thermodynamics. Contributions from the gapless magnons (\ref{k2})
are weighted by the exponentially small probability of being in a
configuration that supports such long-range excitations. For
example, the low-temperature specific heat is governed by a
competition between two distinct contributions: one from the
gapless modes, $C_{1}\sim T^{d/2}e^{-\mathcal{S}_{0}\mathcal{N}}$
(where $d$ is the lattice dimension and $\mathcal{S}_{0}$ is the
residual entropy per spin), and another from the gapped
excitations of the dominant finite clusters, $C_{2}\sim (\Delta
/T)^{2}e^{-\Delta /T}$ (with $\Delta$ is a typical cluster gap).

Comparing these terms shows that $C_{1}$ dominates only at
extremely low temperatures, $T<\Delta
/(\mathcal{S}_{0}\mathcal{N)}$. In the thermodynamic limit $N\to
\infty$, this temperature scale vanishes. Therefore, despite the
formal presence of gapless excitations in the spectrum, the
low-temperature thermodynamics of the system is effectively
governed by gapped, local excitations, resembling that of a
paramagnet with a finite energy gap.

\section{Distorted star model on 2D and 3D lattices}

\begin{figure}[tbp]
\includegraphics[width=3in,angle=0]{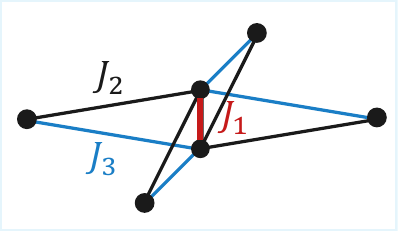}
\caption{Example of distorted star. Exchanges between dimer spins
$J_{1}$ are shown by red lines, the exchanges of monomer spins
with dimer spins, $J_{2}$ and $J_{3}$, are shown by black and blue
lines. } \label{Fig_distorted1}
\end{figure}

We now extend the connected-star model to include a structural
distortion, where each monomer spin is coupled to the two spins of
a neighboring dimer via two distinct exchange integrals, $J_{2}$
and $J_{3}$ (see Fig. \ref{Fig_distorted1}). To realize
macroscopic ground-state degeneracy, the condition for each
triangular plaquette - formed by two dimer spins and one monomer
spin - must be satisfied: its ground-state manifold must include
both a quartet ($S=\frac{3}{2}$) and one of its two doublets
states ($S=\frac{q}{2}$). For a single triangle, this requires
\begin{equation}
\frac{1}{J_{1}}+\frac{1}{J_{2}}+\frac{1}{J_{3}}=0  \label{cond1}
\end{equation}

For the star containing $z$ edge-shared triangles, the conditions
(\ref{cond1}) transforms to
\begin{equation}
\frac{z}{J_{1}}+\frac{1}{J_{2}}+\frac{1}{J_{3}}=0  \label{cond2}
\end{equation}

Throughout, we assume that black exchanges $J_{2}<0$ is
ferromagnetic and $|J_{2}|>|J_{3}|$.

Multiple inequivalent distributions of $J_{2}$ and $J_{3}$ bonds
are possible on a given lattice. To compute the ground-state
degeneracy for a specific arrangement, we apply the method
proposed in Ref. \cite{DKV2D}: set $J_{1}=J_{3}=0$, leaving only
the ferromagnetic $J_{2}$ bonds. The system then effectively
decouples into numerous independent ferromagnetic clusters. The
product of the multiplet degeneracies of all clusters gives the
total ground-state degeneracy, which remains exact for any $J_{1}$
and $J_{3}$ satisfying Eq.\ref{cond2}.

\begin{figure}[tbp]
\includegraphics[width=4in,angle=0]{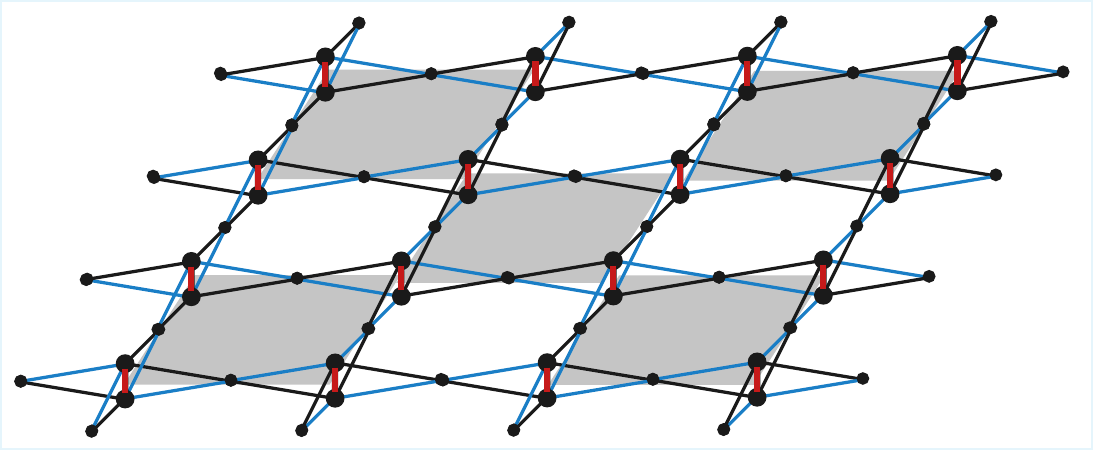}
\caption{Distorted star model on square lattice. Shaded squares
denote trapping cells.} \label{Fig_distorted2}
\end{figure}

For the square lattice, the distribution shown in
Fig.\ref{Fig_distorted2} maximizes the degeneracy. In this
configuration, the $J_{2}$ (black) bonds form disjoint squares,
each enclosing eight spins-$\frac{1}{2}$. An isolated square has a
ferromagnetic ground state degeneracy of $2S+1=9$ (for total spin
$S=4$). Since the squares are decoupled when $J_{1}=J_{3}=0$, the
total degeneracy is $W=9^{N/8}$, where $N$ is the number of
lattice sites (stars). The corresponding residual entropy per spin
is $S_{0}=\frac{1}{8}\ln 9\simeq 0.275$.

\begin{figure}[tbp]
\includegraphics[width=4in,angle=0]{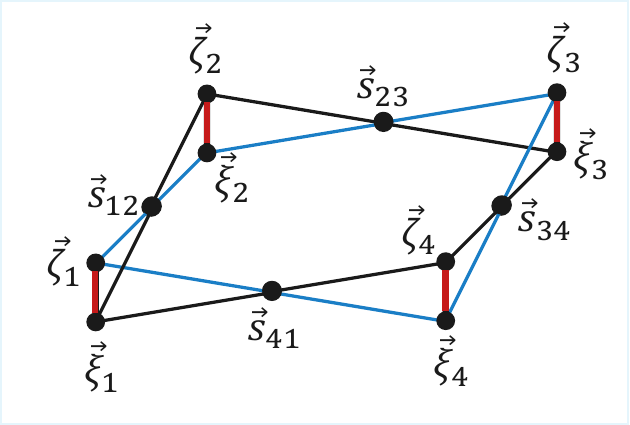}
\caption{Trapping cell for distorted-star model.}
\label{Fig_trapping}
\end{figure}

This result can be verified by constructing the exact localized
multi-magnon ground states within a single square trapping cell,
shown in Fig.\ref{Fig_trapping}. The one-magnon state is
\begin{equation}
\hat{\varphi}_{1}=\left(
s_{12}^{-}+s_{23}^{-}+s_{34}^{-}+s_{41}^{-}+\sigma _{1}^{-}+\sigma
_{2}^{-}+\sigma _{3}^{-}+\sigma _{4}^{-}\right) \left\vert
F\right\rangle   \label{1}
\end{equation}%
where
\begin{equation}
\sigma _{i}^{-}=\frac{\xi _{i}^{-}+J\zeta _{i}^{-}}{1+J}
\label{sigma}
\end{equation}
and $\xi _{i}^{-},\zeta _{i}^{-}$ are lowering operators of dimer
spins and $J=J_3/J_2$ ($J_1$ is defined by Eq.(\ref{cond2})).

Higher multi-magnon states are generated recursively:
\begin{equation}
\hat{\varphi}_{2}=\hat{\varphi}_{1}^{2}-\sum \sigma
_{i}^{-2}\left\vert F\right\rangle   \label{2}
\end{equation}%
and, generally, $k$-magnon localized state in the trapping cell
($k=1,2,...8)$ is
\begin{equation}
\hat{\varphi}_{k}=\hat{\varphi}_{1}^{k-2}\left( \hat{\varphi}_{1}^{2}-\frac{%
k\left( k-1\right) }{2}\sum \sigma _{i}^{-2}\right) \left\vert
F\right\rangle   \label{k}
\end{equation}%
so that, the maximal possible eight-magnon state is
\begin{equation}
\hat{\varphi}_{8}=s_{12}^{-}s_{23}^{-}s_{34}^{-}s_{41}^{-}\sigma
_{1}^{-}\sigma _{2}^{-}\sigma _{3}^{-}\sigma _{4}^{-}\left\vert
F\right\rangle   \label{8}
\end{equation}

Thus, a single square supports nine exact states (with magnon
numbers $0,1,2\ldots 8$).

Crucially, the product of exact localized magnon states from
neighboring squares is not itself an exact eigenstate of the full
Hamiltonian. However, as shown in \cite{DKV2D}, an exact
eigenstate can be formed by adding a specific correction term:
\begin{equation}
\hat{\varphi}_{k}^{(1)}\hat{\varphi}_{m}^{(2)}\left\vert
F\right\rangle
+2J\xi _{12}^{-}\zeta _{12}^{-}\hat{\varphi}_{k-1}^{(1)}\hat{\varphi}%
_{m-1}^{(2)}\left\vert F\right\rangle   \label{corr-term}
\end{equation}%
where $\xi _{12}^{-},\zeta _{12}^{-}$ are spin operators on the
shared dimer between two neighboring trapping cells $1$ and $2$.
This construction demonstrates that there is no hard-core
repulsion preventing magnons from occupying adjacent trapping
cells; the exact many-body ground states are products of the
localized multi-magnon states, appropriately corrected along
shared dimers. Consequently, the total ground-state manifold is
equivalent to that of $\frac{N}{2}$ non-interacting spins of
length $s=4$, confirming $W=9^{N/8}$ and the corresponding
residual entropy. Therefore, at low temperatures the system mimics
a paramagnetic crystal of non-interacting spins $s=4$.

A similar construction for the cubic lattice yields disjoint cubes
formed by black $J_2$ bonds, each containing 20 spins-1/2 and a
ferromagnetic degeneracy of 21. This gives the ground state
degeneracy of the total system $W=21^{N/20}$, which produces the
residual entropy $S_{0}=\frac{1}{20}\ln 21\simeq 0.152$.

Both values are significantly lower than the residual entropies of
the ideal (undistorted) star models (see Table 1), indicating that
distortion reduces the ground state degeneracy. These analytical
results are confirmed by exact diagonalization of finite systems.

As noted previously, in the limit $J_{1}=J_{3}=0$, the system 
decouples into non-interacting ferromagnetic clusters (trapping cells), 
each containing eight spins $\frac{1}{2}$. The lowest excitation above 
the macroscopically degenerate ground state in this limit corresponds 
to the first excited state of an individual cluster, which is finite. 
Consequently, the spectrum is gapped at $J_{1}=J_{3}=0$.

The full one-magnon excitation spectrum for the distorted model, 
calculated with general couplings satisfying Eq. (\ref{cond2}), 
remains gapped for all allowed values of $J_1,J_2,J_3$ . 
Together, these results confirm the existence of a finite energy gap 
$\Delta E$ in the spectrum of the distorted-star model. 
Consequently, at temperatures $T<\Delta E$, the low-field magnetization 
and susceptibility correspond to a paramagnet of non-interacting spins $s=4$:
$m=\frac{11}{3}\frac{h}{T}$ and $\chi=\frac{11}{3}T$ for $h\ll T$.

We have focused here on a specific, optimal arrangement of
exchanges $J_{2}$ and $J_{3}$ for the square lattice. The
construction generalizes directly to other lattices, subject only
to the constraint that every monomer spin connects to its
neighboring dimer via two distinct exchanges satisfying
Eq.(\ref{cond2}). Different distributions of $J_{2}$ and $J_{3}$
will yield different, typically smaller, ground-state
degeneracies.

\section{Summary}

In this work, we have explored a class of highly frustrated
spin-1/2 Heisenberg models on two- and three-dimensional lattices
constructed from connected star units, where each star comprises
edge-sharing triangles, featuring antiferromagnetic couplings
along shared edges and ferromagnetic couplings on the others. Our
analysis reveals that these architectures support exceptionally
high ground-state degeneracy and corresponding residual entropy.

For the ideal (undistorted) star model at the quantum critical
point $J_{1}=2J_{2}$, we established an exact mapping of the
ground state manifold onto a site percolation problem on the Lieb
lattice. This powerful correspondence allowed us to calculate the
exponential degeneracy $W(N)\sim e^{\alpha N}$ and the
thermodynamic residual entropy $\mathcal{S}_{0}$ for square,
triangular, honeycomb, and cubic lattices. The obtained entropy is
not only systematically larger than that of related
diamond-decorated lattices but also remarkably robust, maintaining
a high value (exceeding 60\% of the maximal value $\ln 2$)
independent of the lattice coordination number $z$. Despite the
formal presence of gapless, quadratic ferromagnetic excitations
$E(k)\sim k^2$ in the one-magnon spectrum, the low-temperature
thermodynamics is dominated by the exponentially numerous, gapped
excitations of small, finite ferromagnetic clusters. Consequently,
observables like the specific heat behave as in a gapped system,
with the contribution from the gapless modes being exponentially
suppressed by the large residual entropy. This reflects a
remarkable fact: while the spectrum is gapless, the thermodynamic
weight of the corresponding low-energy states is negligible in the
thermodynamic limit. The zero-temperature magnetization vanishes
as $m\sim N^{-1/2}$, confirming the absence of long-range order
due to the statistical dominance of configurations fragmented
below the geometric percolation threshold.

For the distorted star model, macroscopic degeneracy arises under
the generalized condition $z/J_1 + 1/J_2 + 1/J_3 = 0$. In the
optimal bond arrangement for the square lattice, this leads to a
ground state manifold equivalent to a system of non-interacting
high-spin clusters, yielding a degeneracy $W=9^{N/8}$ and entropy
$\mathcal{S}_{0}=0.275$. The construction generalizes to three
dimensions, giving $W=21^{N/20}$ and $\mathcal{S}_{0}=0.152$ for
the cubic lattice. While these values are lower than those for the
ideal star model, the degeneracy remains exponential and the
excitation spectrum is gapped, leading to conventional
low-temperature paramagnetic behavior.

In summary, we have shown that the architecture of connected-star
lattices provides a design principle for quantum magnets with high
residual entropy. The macroscopic degeneracy at quantum critical
point identifies these systems as promising candidates for
magnetocaloric cooling at ultra-low temperatures and as `entropic
levers' for quantum thermal machines. Our results establish a
framework for engineering high-entropy magnetic materials for
low-temperature applications.

\bibliography{Lieb_Lattice}

\end{document}